\documentclass[twocolumn,showpacs,fleqn,nobibnotes]{revtex4}

\usepackage{amsmath}
\usepackage{graphicx}
\usepackage{float}
\usepackage{subfigure}

\def\lsim{\raise0.3ex\hbox{$<$\kern-0.75em\raise-1.1ex\hbox{$\sim$}}}
\def\gsim{\raise0.3ex\hbox{$>$\kern-0.75em\raise-1.1ex\hbox{$\sim$}}}

\newcommand{\rr}{\mbox{\boldmath $r$}}

\begin{document}

\title{Coherent photon-hadron interactions in $pA$ collisions: Small-$x$ physics after HERA}
\author{V.P. Gon\c{c}alves
$^{a}$
and M.V.T. Machado $^{b,c}$ }

\affiliation{$^a$ Instituto de F\'{\i}sica e Matem\'atica, Universidade Federal de
Pelotas\\
Caixa Postal 354, CEP 96010-090, Pelotas, RS, Brazil\\
$^b$ Universidade Estadual do Rio Grande do Sul (UERGS). Unidade de Bento Gon\c{c}alves \\ Rua Benjamin Constant, 229. Bento Gon\c{c}alves. CEP 95700-000, Brazil\\
$^c$  High Energy Physics Phenomenology Group, GFPAE  IF-UFRGS \\
Caixa Postal 15051, CEP 91501-970, Porto Alegre, RS, Brazil
}

\begin{abstract}
In this letter we study the photoproduction of heavy quarks and vector mesons in the coherent proton-nucleus ($pA$) interactions for  RHIC and LHC energies and  analyze if these processes can be used to determine the QCD dynamics at high energies. The integrated cross section and rapidity distribution are estimated using the Color Glass Condensate (CGC) formalism. A comparison with the linear dynamics predictions is also presented. Our results indicate that the nonlinear dynamics can be proven in those reactions, which are well suited for studing saturation effects.

\end{abstract}

\maketitle

\section{Introduction} 

Over the last decade much progress has been realized towards understanding the QCD dynamics at high energies. The successful operation of the DESY $ep$ collider HERA has opened a new era of  experimental and theoretical investigation into the deep structure of the proton and, in general, of hadronic interactions. Some of the  most important observations are the striking rise of the proton structure function $F_2(x,Q^2)$ for small values of the Bjorken variable $x$ ($< 10^{-2}$),  the large contribution of diffractive processes in this kinematical range and the geometric scaling (For a recent review, see
e.g. Ref. \cite{Abra}). Theoretically, at small $x$, due to the large gluon density,   we expect the transition of the
regime described by the linear dynamics, where only the parton emissions are considered,
for a new regime where the physical process of recombination of partons
become important in the parton cascade and the evolution is given by a
nonlinear evolution equation.  This regime is characterized by the limitation
on the maximum phase-space parton density that can be reached in the hadron
wavefunction (parton saturation), with the transition being  specified  by a typical scale, which is energy dependent and is called saturation scale $Q_{\mathrm{sat}}$ (For recent reviews see Ref. \cite{hdqcd}).
 Although the HERA experimental  results have a natural interpretation in terms of the saturation physics, due to the kinematical limitations of the experiment none of these phenomena can be taken as a conclusive evidence for a new regime of the QCD dynamics. Currently, the HERA II run has obtained much more precise experimental data, but they are limited at center of mass energies smaller than  300 GeV, where deviations between linear and saturation predictions are small. Consequently, our understanding of the correct QCD dynamics at high energies is still an open question. As HERA will stop its operations, in principle,  in 2007, and the QCD dynamics must be known as precisely as possible in order to maximize the discovery potential for new physics at the next generation of colliders, the study of alternatives which could constrain the QCD dynamics is timely and necessary. 

In this letter we study the   photoproduction of heavy quarks and vector mesons in the coherent proton-nucleus ($pA$) interactions at RHIC and LHC energies and  analyze if these processes can be used to determine the QCD dynamics at high energies (For similar studies on $pp$ and $AA$ collisions, see Refs. \cite{vicmag_upcs,vicmag_prd}). The main advantage of using colliding hadron and nuclear beams for studying photon induced interactions is the high equivalent photon energies and luminosities that can be achieved at existing and future accelerators (For a recent discussion see Ref. \cite{frank}).

\section{Coherent $pA$ interactions} 

Lets consider the proton-nucleus interaction at large impact parameter ($b > R_p + R_A$) and at ultrarelativistic energies. In this regime we expect the electromagnetic interaction to be dominant.  When we consider the electromagnetic field associated to the proton and ion, we have that due to the coherent action of all protons in the nucleus, the electromagnetic field surrounding the ion is very larger than the proton one. This result can be easily understood if we use the Weisz\"acker-Williams method to calculate the equivalent flux of photons  from a charge
$Z$ nucleus a distance $b$ away, which is given by (For recent reviews see Ref.  \cite{upcs})
\begin{eqnarray}
\frac{d^3N_{\gamma}\,(\omega,\,b^2)}{d\omega\,d^2b}= \frac{Z^2\alpha_{em}\eta^2}{\pi^2 \,\omega\, b^2}\, \left[K_1^2\,(\eta) + \frac{1}{\gamma_L^2}\,K_0^2\,(\eta) \right] \,
\label{fluxunint}
\end{eqnarray}
where $\omega$ is the photon energy,  $\gamma_L$ is the Lorentz boost  of a single beam and $\eta
= \omega b/\gamma_L$; $K_0(\eta)$ and  $K_1(\eta)$ are the
modified Bessel functions. From the above expression we have that photon spectrum of a nucleus with charge $Z$ is proportional to $Z^2$.
Due to asymmetry in the collision, with the ion being likely the photon emitter, we have that the photon direction is known, which will implicate an asymmetry in the rapidity distribution (see below).    
 The coherence condition limits the photon virtuality to very low values $(Q^2 \le 1/R_A^2)$, which implies that for most purposes, these can be considered as real. 
Therefore, coherent $pA$ collisions can be used in order to study photon-proton interactions.

 The requirement that  photoproduction
is not accompanied by hadronic interaction (ultraperipheral
collision) can be done by restricting the impact parameter $b$  to
be larger than the sum of the proton and  the nuclear radius.
Therefore, the total photon flux interacting with the target
nucleus is given by Eq. (\ref{fluxunint}) integrated over the
transverse area of the target for all impact parameters subject to
the constraint that the proton and the nucleus  do not interact hadronically.
An analytic approximation for $pA$ collisions can be obtained
using as integration limit $b> R_p + R_A$, producing \cite{upcs}
\begin{eqnarray}
\frac{dN_{\gamma}\,(\omega)}{d\omega}= \frac{2\,Z^2\alpha_{em}}{\pi\,\omega}\, \left[\bar{\eta}\,K_0\,(\bar{\eta})\, K_1\,(\bar{\eta})+ \frac{\bar{\eta}^2}{2}\,{\cal{U}}(\bar{\eta}) \right]\,
\label{fluxint}
\end{eqnarray}
where $\bar{\eta}=\omega\,(R_p + R_A)/\gamma_L$ and  ${\cal{U}}(\bar{\eta}) = K_1^2\,(\bar{\eta})-  K_0^2\,(\bar{\eta})$. 
The cross section for the photoproduction of a final state $X$ in a coherent  $pA$
collisions will be  given by,
\begin{eqnarray}
\sigma (Ap \rightarrow X Y)\, = \int \limits_{\omega_{min}}^{\infty} d\omega \frac{dN_{\gamma}(\omega)}{d\omega}\,\sigma_{\gamma p \rightarrow X Y} \left(W_{\gamma p}^2\right)\,,
\label{sigAA}
\end{eqnarray}
where $\omega_{min}=M_{X}^2/4\gamma_L m_p$, $W_{\gamma p}^2=2\,\omega\sqrt{S_{\mathrm{NN}}}$  and
$\sqrt{S_{\mathrm{NN}}}$ is  the c.m.s energy of the
proton-nucleus system. Considering $p Pb \, (Ar)$ collisions at LHC, the Lorentz factor  is
$\gamma_L=4690 \,(5000)$, giving the maximum c.m.s. $\gamma N$ energy
$W_{\gamma p} \approx 1500 \,(2130)$ GeV. Therefore, while studies of photoproduction at HERA are limited to photon-proton center of mass energies of about 200 GeV, photon-proton interactions at  LHC can reach one order of magnitude higher on energy. Consequently, studies of $\gamma p$ interactions at LHC could provide valuable information on the QCD dynamics at high energies.  
In this work we consider that the produced state $X$ represents a  $Q\overline{Q}$ pair or a vector meson $V$. Since photon emission is coherent over the entire nucleus and the photon is colorless we expect that the events to be characterized by one  (heavy quark production, with $Y$ being the remaining  of the proton) or two  (vector meson production, with $Y = p$) rapidity gaps.

\section{QCD dynamics at high energies} 

The photon-hadron interaction at high energy (small $x$) is usually described in the infinite momentum frame  of the hadron in terms of the scattering of the photon off a sea quark, which is typically emitted  by the small-$x$ gluons in the proton. However, in order to disentangle the small-$x$ dynamics of the hadron wavefunction, it is more adequate to consider the photon-hadron scattering in the dipole frame, in which most of the energy is
carried by the hadron, while the  photon  has
just enough energy to dissociate into a quark-antiquark pair
before the scattering. In this representation the probing
projectile fluctuates into a
quark-antiquark pair (a dipole) with transverse separation
$\rr$ long after the interaction, which then
scatters off the proton \cite{nik}. The main motivation to use this color dipole approach, is that it gives a simple unified picture of inclusive and diffractive processes. In particular,  in this approach  the heavy quark photoproduction cross section  reads as,
\begin{eqnarray}
\sigma\,(\gamma p \rightarrow Q\overline{Q}X)= \sum_{h, \bar{h}}
\int dz\, d^2\rr \,\Psi^\gamma_{h, \bar{h}}\,\sigma_{dip}(x,\rr) \,\Psi^{\gamma *}_{h, \bar{h}} \,
\end{eqnarray} 
where 
$\Psi^{\gamma}_{h, \bar{h}}(z,\,\rr)$  is the light-cone wavefunction  of the photon \cite{nik}. 
   The quark and antiquark helicities are labeled by $h$ and $\bar{h}$. The variable $\rr$ defines the relative transverse
separation of the pair (dipole) and $z$ $(1-z)$ is the
longitudinal momentum fractions of the quark (antiquark). The basic
blocks are the photon wavefunction, $\Psi^{\gamma}$  and the dipole-target  cross
section, $\sigma_{dip}$. For photoproduction we have that longitudinal piece does not contribute, since $|\Psi_{L}|^2\propto Q^2$, and the total cross section is computed introducing the appropriated mass and charge of the charm or  bottom quark.

Similarly, in the dipole picture the
imaginary part of the amplitude for vector meson  production at zero momentum
transfer reads  as (See e.g. Refs. \cite{nik,victor_magno_mesons})
\begin{eqnarray}
{\cal I}m \, {\cal A}\, (\gamma p \rightarrow Vp)  = \sum_{h, \bar{h}}
\int dz\, d^2\rr \,\Psi^\gamma_{h, \bar{h}}\,\sigma_{dip}(\tilde{x},\rr) \, \Psi^{V*}_{h, \bar{h}} \, ,
\label{sigmatot}
\end{eqnarray}
where  $\Psi^{V}_{h,
  \bar{h}}(z,\,\rr)$  is the light-cone wavefunction  of the vector meson. 
The total
cross section for vector meson photoproduction is given by 
\begin{eqnarray}
\sigma\, (\gamma p \rightarrow Vp) = \frac{[{\cal I}m \, {\cal A}(s,\,t=0)]^2}{16\pi\,B_V}\,(1+\beta^{2}) \;
\label{totalcs}
\end{eqnarray}
where $\beta$ is the ratio of real to imaginary part of the
amplitude and $B_V$ labels the slope parameter. The values considered
for the slope parameter are taken from the  parameterization used
in Ref. \cite{victor_magno_mesons}.  

\begin{table}[t]
\begin{center}
\begin{tabular} {||c|c|c|c||}
\hline
\hline
& $X$   & {\bf COLLINEAR } & {\bf CGC}  \\
\hline
\hline
 {\bf LHC} & $c\bar{c}$ &  17 mb ($10^{10}$) & 5 mb ($1\cdot 10^9$) \\
\hline
&  $b\bar{b}$ &  155 $\mu$b ($10^8$) &  81 $\mu$b ($6\cdot 10^7$) \\
\hline 
&  $\rho $ &  --- &  14 mb ($1\cdot 10^{10}$) \\
\hline
&  $J/\Psi $ &  ---  &  95 $\mu$b ($7\cdot 10^{7}$) \\
\hline
\hline
\end{tabular}
\end{center}
\caption{\it The integrated cross section (event rates/month) for the photoproduction of heavy quarks and vector mesons in $pA$ collisions at LHC (see text).}
\label{tabhq}
\end{table}

We have that the total cross sections for vector meson and heavy quark production in dipole approach are strongly  dependent on the dipole-hadron cross section $\sigma_{dip}$, which   contains all
information about the target and the strong interaction physics.
In the Color Glass Condensate (CGC)  formalism \cite{CGC,BAL,WEIGERT}, $\sigma_{dip}$ can be
computed in the eikonal approximation, resulting
\begin{eqnarray}
\sigma_{dip} (x,\rr)=2 \int d^2 b_{\perp} \,\left[
1-\mathrm{S}\,(x,\rr,b_{\perp})\right]\,\,,
\end{eqnarray}
where $S$ is the $S$-matrix element which encodes all the
information about the hadronic scattering, and thus about the
non-linear and quantum effects in the hadron wave function. The
function $S$ can be obtained by solving an appropriate evolution
equation in the rapidity $y\equiv \ln (1/x)$. The main properties
of $S$ are: (a) for the interaction of a small dipole ($\rr
\ll 1/Q_{\mathrm{sat}}$), $S(\rr) \approx 1$, which characterizes that
this system is weakly interacting; (b) for a large dipole
($\rr \gg 1/Q_{\mathrm{sat}}$), the system is strongly absorbed which
implies $S(\rr) \ll 1$.  This property is associate to the
large density of saturated gluons in the hadron wave function. In
our analysis we will  consider the  phenomenological
saturation model proposed in Ref. \cite{IIM} which encodes the
main properties of the saturation approaches, with the dipole cross section  parameterized as follows
\begin{eqnarray}
\sigma_{dip}^{\mathrm{CGC}}\,(x,\rr) =\sigma_0\,\left\{ \begin{array}{ll} 
{\mathcal N}_0 \left(\frac{\bar{\tau}^2}{4}\right)^{\gamma_{\mathrm{eff}}\,(x,\,r)}\,, & \mbox{for $\bar{\tau} \le 2$}\,, \nonumber \\
 1 - \exp \left[ -a\,\ln^2\,(b\,\bar{\tau}) \right]\,,  & \mbox{for $\bar{\tau}  > 2$}\,, 
\end{array} \right.
\label{CGCfit}
\end{eqnarray}
where $\bar{\tau}=\rr Q_{\mathrm{sat}}(x)$ and the expression for $\bar{\tau} > 2$  (saturation region)   has the correct functional
form, as obtained  from the theory of the Color Glass Condensate (CGC) \cite{CGC}. For the color transparency region near saturation border ($\bar{\tau} \le 2$), the behavior is driven by the effective anomalous dimension $\gamma_{\mathrm{eff}}\, (x,\,r)= \gamma_{\mathrm{sat}} + \frac{\ln (2/\tilde{\tau})}{\kappa \,\lambda \,y}$, where $\gamma_{\mathrm{sat}}=0.63$ is the LO BFKL anomalous dimension at saturation limit. Hereafter, we label this model  by CGC.

\begin{figure}[t]
\includegraphics[scale=0.47]{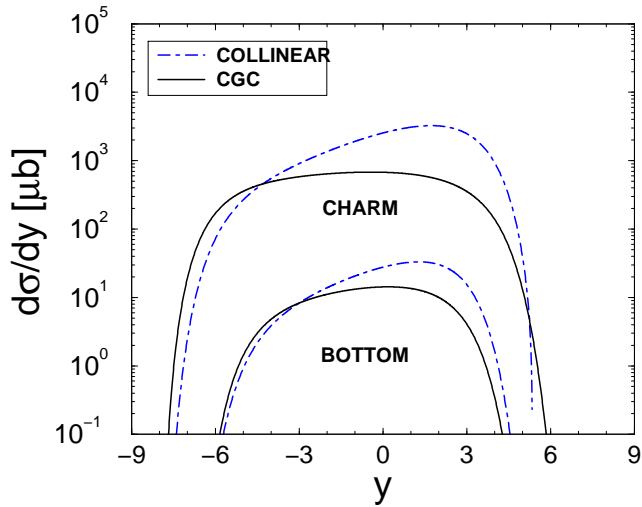}
\caption{(Color online) Rapidity distribution for heavy quark photoproduction on $pA$ reactions for LHC energy (see text).}
\label{fig:1}
\end{figure}

\section{Results}
 
The distribution on rapidity $y$ of the produced final state can be directly computed from Eq. (\ref{sigAA}), by using its  relation with the photon energy $\omega$, i.e. $y\propto \ln \, (2 \omega/m_X)$.  Explicitly, the rapidity distribution is written down as,
\begin{eqnarray}
\frac{d\sigma \,\left[A+p \rightarrow  X + Y \right]}{dy} = \omega \frac{dN_{\gamma} (\omega )}{d\omega }\,\sigma_{\gamma p \rightarrow X Y}\left(\omega \right).
\label{dsigdy}
\end{eqnarray}
Consequently, given the photon flux, the rapidity distribution is thus a direct measure of the photoproduction cross section for a given energy.
In Fig. \ref{fig:1} we present our results for the heavy quark photoproduction at LHC energies.  For comparison, we also present the predictions from the linear dynamics, denoted collinear hereafter, which is calculated assuming that the collinear factorization is valid  and that the gluon distribution can be described by the GRV98 parametrization (For more details see Ref. \cite{vicmag_prd}). In Tab. \ref{tabhq} one presents the correspondent integrated cross sections (event rates), using a luminosity of ${\cal L}_{\mathrm{pPb}}=7.4\times 10^{29}$ cm$^2$s$^{-1}$. We have verified for completeness that for RHIC energy the difference between the predictions for the rapidity distribution is small, which is expect due to small value of the photon-proton center of  mass energy. However, at LHC we can observe a large difference between the predictions and having  high rates. For instance, for charm quark CGC gives a cross section a factor 3 lower than collinear and a factor 2 for bottom. This deviation holds even in case of experimental cuts on rapidity. Therefore, photoproduction of heavy quarks should provide a feasible and clear measurement of the underlying QCD dynamics at high energies. Since RHIC is obtaining data for $d Au$ interactions and LHC is to be commissioned, these processes could be analyzed in the next years. The advantages are a clear final state (rapidity gap and low momenta particles) and no competing effect of dense nuclear environment if compared with hadroproduction.

\begin{figure}[t]
\includegraphics[scale=0.47]{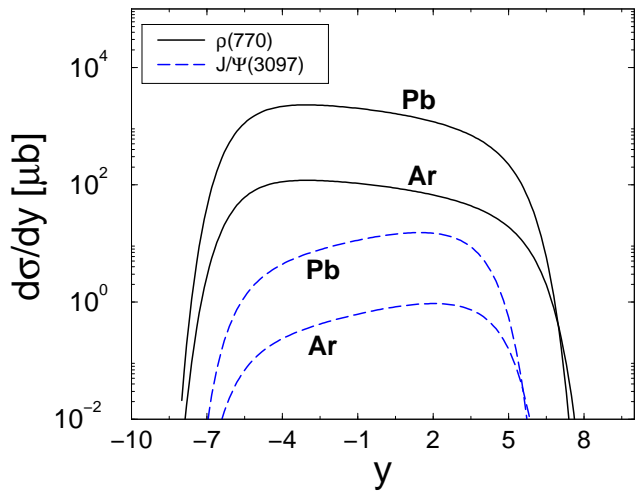}
\caption{(Color online) Rapidity distribution for vector meson photoproduction on $pA$ reactions at LHC energy (see text).}
\label{fig:2}
\end{figure}
The present results can be directly compared with those obtained in Ref. \cite{klein_vogt}, which use only collinear approach. The orders of magnitude are similar for RHIC and LHC, with main deviations coming from  different choices for the  gluon pdfs and distinct factorization scale. Our estimates considering CGC (collinear) for RHIC are 142 (110) $\mu$b for charm and 0.15 (0.10) $\mu$b for bottom.  The rapidity distributions can be also contrasted with estimations in Refs. \cite{vicmag_upcs,vicmag_prd}, where the heavy quark photoproduction in $AA$ and $pp$ collisions have been computed. To see what is difference in the order of magnitude among them let us perform a few parametric estimates. Roughly, the photon flux on nuclei is approximately $Z^2$ the flux on proton, $dN_{\gamma}^A/d\omega \propto Z^2\, dN_{\gamma}^p/d\omega$. Moreover, for heavy quarks have been not verified large nuclear shadowing \cite{vicmag_upcs} such that $\sigma_{\gamma A}\approx A\,\sigma_{\gamma p}$. The ratio between $pA$ production and $pp$ (or $AA$) can be estimated as $R_{pA/pp(AA)}=\frac{d\sigma_{pA}/dy}{d\sigma_{pp(AA)}/dy}$. Therefore, using Eq. (\ref{dsigdy}), one obtains $R_{pA/pp}\propto Z^2$ and the enhancement reaches a factor $10^4$ for heavy nuclei in the comparison between photoproduction on $pA$ and in energetic protons. On the other hand, one obtains $R_{pA/AA}\propto A^{-1}$ and then $pA$ is suppressed in relation to $AA$ by a factor $A$. However, the larger $pA$ luminosity, which is two order of magnitude higher than for $AA$, counteracts this suppression for the event rates.

\begin{figure}[t]
\includegraphics[scale=0.47]{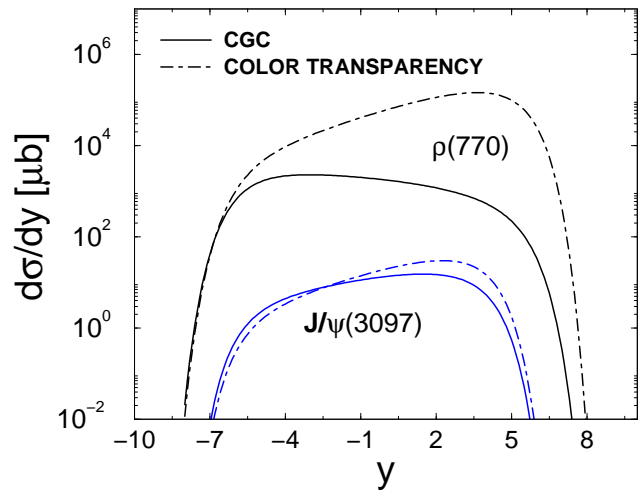}
\caption{(Color online) Comparison between CGC and color transparency estimates (see text).}\label{fig:3}
\end{figure}

Concerning coherent meson production, in Fig. \ref{fig:2} one presents predictions for the rapidity distribution on $pA$ collisions at LHC energy considering both light ($Ar$) and heavy ($Pb$) ions. These theoretical predictions are original in the literature. The corresponding integrated cross sections (event rates) are shown in Tab. \ref{tabhq}, having  high rates. For completeness, we have computed the estimation for typical light meson ($\rho$) and a heavy one ($J/\Psi$). For RHIC, the cross section are 13 mb (0.8 $\mu$b) for $\rho$ ($J/\Psi$) production. These results can be contrasted with those obtained in the Refs. \cite{vicmag_upcs,vicmag_prd,klein_nis_prl}. We can do a similar estimative of the ratio  as for heavy quarks considering that for light mesons $\sigma_{\gamma A}\approx A^{2/3}\,\sigma_{\gamma p}$ and for heavy mesons $\sigma_{\gamma A}\approx A^{4/3}\,\sigma_{\gamma p}$ \cite{vicmag_upcs}. Therefore, an enhancement of $Z^2$ for $pA/pp$  remains but now $R_{pA/AA}\propto A^{-2/3}\, (A^{-4/3})$, respectively.  For Pb ions at LHC this is a factor 35 for $\rho$ and a factor $10^3$ for $J/\Psi$. It should be noticed that the coherent production of mesons is currently measured at RHIC for the $AA$ case. Therefore, the present estimation could be tested in $dAu$ collisions with a good experimental feasibility. For comparison, we also present in Fig. \ref{fig:3} the prediction from the  color transparency limit of the CGC model, which is obtained assuming that $\sigma_{dip}^{\mathrm{CGC}} \propto (\rr Q_{\mathrm{sat}})^2$ is given in all kinematic range by its value for $\bar{\tau} < 1$. It is equivalent to disregard the saturation physics, which allows to estimate the importance of the CGC physics in the process. We have that the predictions for $J/\Psi$ production are somewhat similar, which is expected, since the heavy vector meson production is dominated by small pair separations, where the saturation physics does not contribute significantly. On the other hand, the photoproduction of $\rho$ mesons is dominated by physics below saturation scale. The difference is huge, reaching 3 orders of magnitude in rapidity distribution, which demonstrate the importance of the saturation physics on this process.

Finally, lets comment on the uncertainties and model dependence for the results  presented here.  For the heavy quarks (HQ) production, the main theoretical uncertainties are the higher order corrections in perturbative expansion and the quark masses. Within the perturbative QCD collinear formalism, the higher order corrections for HQ photoproduction are quite small (a few percents) in contrast to the hadroproduction case and they can be taken into account by a suitable choice for the quark mass \cite{klein_vogt,mag_cris}. The color dipole formalism, which we use in our analysis, contains contributions of higher orders by definition as it is equivalent to the semihard factorization (or $k_t$-factorization) approach in leading logarithm approximation. The latter is well known to include higher order perturbative QCD diagrams in its formulation. Concerning the DGLAP results (collinear formalism), the uncertainties for different choices of the gluon distribution are small as  it is probed at a relatively large scale $\mu^2 = 4m_Q^2$ (approximately 9 GeV$^2$ for charm and 81 GeV$^2$ for bottom).  Those uncertainties are known to take place at low virtualities $Q^2=$ 1-3 GeV$^2$. Moreover, our results are an extrapolation of the photoproduction cross section for energies beyond  energies of current accelerator regime using the physical parameters which describe correctly both the low energy (fixed target) and the HERA data points. Therefore, the present estimates should be reliable. Concerning vector meson production, in general it is described by quite distinct approaches for light and heavy mesons. In the light meson case, the old vector meson dominance (VMD) model is often adopted \cite{klein_prc}, whereas for the heavy mesons a leading logarithmic approximation of the collinear approach is considered \cite{strikman_vec}. In this case, the hard QCD scale is given by the vector meson mass. We have used the unique theoretical formalism available describing simultaneously light and heavy vector meson production \cite{victor_magno_mesons}. Once again, the parameters of the model are fixed in order to describe the experimental low energy and DESY-HERA datasets. The transition between light and heavy mesons is dynamically  introduced by parton saturation effects (via saturation scale) in the proton target.

\section{Summary}

The QCD dynamics at high energies is of utmost importance for building a realistic description of $pp/pA/AA$ collisions at LHC. In this limit QCD evolution leads to a system with high gluon density. We have shown if such system there exists at high energies it can be proven in coherent $pA$ collisions at LHC.  We propose two specific final states (heavy quarks and mesons) where the experimental identification could be feasible. It should be noticed, however, that the predictions presented here were based on specific models/assumptions for the process and on recent phenomenological investigations. Therefore, these assumptions should be tested in future $pA$ experimental investigations. 

\vspace{-0.6cm}

\begin{acknowledgments}
This work was  partially financed by the Brazilian funding
agencies CNPq and FAPERGS.
\end{acknowledgments}

\end{document}